\documentclass{aa}
\usepackage{natbib}
\bibpunct{(}{)}{;}{a}{}{,}
\input{psfig}

\def\Real{{\rm I\mathchoice{\kern-0.70mm}{\kern-0.70mm}{\kern-0.65mm}%
  {\kern-0.50mm}R}}

\font \bolditalics = cmmib10
\def\bx#1{\leavevmode\thinspace\hbox{\vrule\vtop{\vbox{\hrule\kern1pt
        \hbox{\vphantom{\tt/}\thinspace{\bf#1}\thinspace}}
      \kern1pt\hrule}\vrule}\thinspace}

\def \vc #1{{\textfont1=\bolditalics \hbox{$\bf#1$}}}

\def\eg{{\bf e}}

\def\thetag{{\vc \theta}}

\def\gammag{{\vc \gamma}}

\def\Cg{{\bf C}}

\def\be{\begin{equation}}
\def\ee{\end{equation}}
\def\ba{\begin{eqnarray}}
\def\ea{\end{eqnarray}}

\def\ltsima{$\; \buildrel < \over \sim \;$}
\def\lsim{\lower.5ex\hbox{\ltsima}}
\def\gtsima{$\; \buildrel > \over \sim \;$}
\def\gsim{\lower.5ex\hbox{\gtsima}}

\begin{document}



   \title{Dealing with systematics in cosmic shear studies: new results from the VIRMOS-Descart Survey\thanks{
Based on observations obtained at the
Canada-France-Hawaii Telescope (CFHT) which is operated by the National
Research Council of Canada (NRCC), the Institut des Sciences de l'Univers
(INSU) of the Centre National de la Recherche Scientifique (CNRS) and
the University of Hawaii (UH)}}

   \author{L. Van Waerbeke$^{1}$, Y. Mellier$^{1,2}$, H. Hoekstra$^{3,4}$}
   \offprints{waerbeke@iap.fr}

   \institute{$^1$ Institut d'Astrophysique de Paris. 98 bis, boulevard
     Arago. 75014 Paris, France. \\
     $^2$ Observatoire de Paris. LERMA. 61, avenue de
     l'Observatoire.  75014 Paris, France.\\
     $^3$ Canadian Institut for Theoretical Astrophysics, 60 St 
     George St., Toronto, M5S 3H8 Ontario, Canada.\\
     $^4$ Dept. of Astronomy \& Astrophysics, University of Toronto, 60 St 
     George St., Toronto, M5S 3H8 Ontario, Canada.
   }

   \markboth{Dealing with systematics in cosmic shear studies}{}



\authorrunning{Van Waerbeke et al.}
\titlerunning{Results from the VIRMOS-Descart Survey}

\abstract{We present a reanalysis of the VIRMOS-Descart weak lensing
data, with a particular focus on different corrections for the
variation of the point spread function anisotropy (PSF) across the
CCDs.  We show that the small scale systematics can be minimised, and
eventually suppressed, using the $B$ mode (curled shear component)
measured in the corrected stars and galaxies. Updated cosmological
constraints are obtained, free of systematics caused by PSF
anisotropy.  To facilitate general use of our results, we provide the
two-points statistics data points with their covariance matrices up
to a scale of one degree.
For the normalisation of the mass power spectrum we obtain
$\sigma_8=(0.83\pm 0.07)\left(\Omega_M/0.3\right)^{-0.49}$. The shape
parameter $\Gamma$ was marginalised over $\Gamma\in[0.1,0.3]$ and the
mean source redshift $z_s$ over $[0.8,1.0]$. The latter is consistent
with recent photometric redshifts obtained for the VIRMOS data and the
preliminary spectroscopic redshifts from the VIRMOS-VVDS survey. The
quoted $68\%$ contour level includes all identified sources of error.  We
discuss the possible sources of residual contamination in this result:
the effect of the non-linear mass power spectrum and remaining issues
concerning the PSF correction. Our result is compared with the first
release of the Wilkinson Microwave Anisotropy Probe data.  It is found
that Cold Dark Matter models with a power law primordial power
spectrum and high matter density $\Omega_M > 0.5$ are excluded at
3-$\sigma$.   }

\maketitle

\section{Introduction}

Weak gravitational lensing by large scale structure (i.e., cosmic
shear) directly probes the matter distribution and complements studies
of the local universe based on the light (galaxy) distribution
alone. Its sensitivity to the dark matter power spectrum and the
geometry of the universe are key ingredients for breaking cosmological
parameters degeneracies associated with other experiments such as
Cosmic Microwave Background \citep{Sp03} and the Type Ia supernovae
\citep{R98,P99}. This powerful application has motivated a first
generation of surveys and the field has made significant strides
forward in the past four years.  A compilation of the first measurements
\citep{BRE00,K00,VW00,WT00} and of the more numerous recent ones,
with discussions of the strength and weaknesses of weak lensing, have been
recently reviewed by a number of
papers \citep[e.g.,][]{VWM03,R03,HOEK03b}.

Cosmic shear studies are now entering the second phase as new surveys,
such as the Canada-France-Hawaii Telescope Legacy Survey 
(CFHTLS)\footnote{{\tt http://www.cfht.hawaii.edu/Science/CFHLS/}} aim
to cover areas of sky that are an order of magnitude larger than
current data sets. These second generation surveys can provide
constraints on a range of cosmological parameters, with a precision
that is comparable to current CMB experiments. However, the success
hinges on the ability to control and correct for observational
systematics. Indeed, there are several technical issues remaining,
concerning the point spread function (PSF) correction, the galaxy
shape analysis and the prediction of the non-linear power spectrum.
Whether this is possible is still debated.

One of the best ways to proceed is to learn from experience obtained
from the first generation surveys. The VIRMOS-Descart survey that will
be discussed below provides an excellent means to do so. It has
already produced several results of cosmological interest \citep{VW02,
HOEK02a, BERN02, P02, PEN03}, but showed a non-zero B-mode signal at
all scales, which has limited its use for precise cosmological
applications. This systematic residual signal is indeed an important
noise contribution. It is therefore timely to understand the cause of
B-mode generation in cosmic shear data and to provide better and more
reliable tools to handle them.

The purpose of this paper is to put the various issues mentioned above
into perspective, and discuss their relevance to future surveys. We do
so by presenting a reanalysis of the VIRMOS-Descart data, taking
advantage of the latest improvements in PSF correction and a better
estimate of the redshift distribution of sources based on recent
spectroscopic data in the VIRMOS-Descart fields obtained by the
VIRMOS-VLT Deep Survey
(VVDS)\footnote{http://www.astrsp-mrs.fr/virmos/vvds.htm}.  Tables
providing results of the best-corrected shear signals as function of
angular scale as well as errors are also given in this paper to help
further joint analyses using complementary cosmological data sets.

This paper is organised as follows: Section 2 defines the notation and
introduces the statistical quantities to be used. Section 3 provides a
summary of the VIRMOS survey. In Section 4 we discuss the impact of
PSF correction on the cosmic shear signal. Sections 5 and 6 show the
new results from the VIRMOS-Descart survey.

\section{Theory}

For self-consistency and in order to define the notations, we briefly
outline the revelant quantities used in this paper. More detailed
calculations and explanations can be found in recent reviews.
Following the notation in \cite{SVWJK98}, we define the power spectrum
of the convergence $\kappa$ as

\begin{eqnarray}
P_\kappa(k)&=&{9\over 4}\Omega_0^2\int_0^{w_H} {{\rm d}w \over a^2(w)}
P_{3D}\left({k\over f_K(w)};
w\right)\times\nonumber\\
&&\left[ \int_w^{w_H}{\rm d} w' n(w') {f_K(w'-w)\over f_K(w')}\right]^2,
\label{pofkappa}
\end{eqnarray}
where $f_K(w)$ is the comoving angular diameter distance out to a
distance $w$ ($w_H$ is the horizon distance), and $n(w(z))$ is the
redshift distribution of the sources given in Eq.(\ref{zsource}).
$P_{3D}(k)$ is the 3-dimensional mass power spectrum (computed from a
non-linear estimation of the dark matter clustering,
see e.g. \cite{PD96,S03}), and $k$ is the 2-dimensional wave vector
perpendicular to the line-of-sight.  The top-hat shear variance
(computed using a smoothing window of radius $\theta_c$) and the shear
correlation function can be written as

\begin{equation}
\langle\gamma^2\rangle={2\over \pi\theta_c^2} \int_0^\infty~{{\rm d}k\over k} P_\kappa(k)
[J_1(k\theta_c)]^2,
\label{theovariance}
\end{equation}

\begin{equation}
\langle\gamma\gamma\rangle_\theta={1\over 2\pi} \int_0^\infty~{\rm d} k~
 k P_\kappa(k) J_0(k\theta).
\label{theogg}
\end{equation}
These statistics could be derived directly from the ellipticity of the
galaxies, but they do not provide a quantitative analysis of residual
systematics.  A more useful measurement of the signal and the
systematics is obtained from the splitting the signal into its
curl-free ($E$ mode) and curl ($B$ mode) components respectively. This
method has been advocated before to help the measurement of the
intrinsic alignment contamination in the weak lensing signal
\citep{CNPT01a,CNPT01b}, but it turns out to be efficient to measure
the residual systematics from the PSF correction as well \citep{P02}.

The $E$ and $B$ modes derived from the shape of galaxies are
unambiguously defined only for the so-called aperture mass variance
$M_{\rm ap}^2$, which is a weighted shear variance within a cell of
radius $\theta_c$. The cell itself is defined as a compensated filter
$U(\theta)$, such that a constant convergence $\kappa$ gives $M_{\rm
ap}=0$:

\begin{equation}
M_{\rm ap}=\int_{\theta < \theta_c}~{\rm d}^2\thetag \kappa(\thetag)~U(\theta).
\end{equation}
$\langle M_{\rm ap}^2\rangle$ can be calculated directly from the shear
$\gammag$ without the need for a mass reconstruction. It can be
rewritten as a function of the shear if we express
$\gammag=(\gamma_t,\gamma_r)$ in the local frame of the line connecting
the aperture center to the galaxy. $M_{\rm ap}$ can therefore be
expressed as function of $\gamma_t$ only
\citep{ME91,K92}:

\begin{equation} M_{\rm ap}=\int_{\theta < \theta_c}~{\rm d}^2\thetag \gamma_t(\thetag)~Q(\theta),
\label{mapfromshear}
\end{equation}
where the filter $Q(\theta)$ is given from $U(\theta)$:

\begin{equation}
Q(\theta)={2\over \theta^2}\int_0^\theta~{\rm d}\theta'~\theta'~U(\theta')-U(\theta)
\label{Qfct}
\end{equation}
The aperture mass variance is related to the convergence power
spectrum (Eq.\ref{pofkappa}) by:

\begin{equation}
\langle M_{\rm ap}^2\rangle={288\over \pi\theta_c^4} \int_0^\infty~{{\rm d}k\over k^3}
 P_\kappa(k) [J_4(k\theta_c)]^2.
\label{theomap}
\end{equation}

The $B$-mode is obtained by replacing $\gamma_t$ with $\gamma_r$ in
Eq.(\ref{mapfromshear}), and the $B$ mode variance is denoted $\langle
M_{\rm ap}^2\rangle_\perp$. The choice of $U$ is arbitrary at this
point, provided it is a compensated filter. In this paper, $U(\theta)$
is chosen as in \cite{SVWJK98},

\begin{equation}
U(\theta)={9\over \pi \theta_c^2} \left(1-{\theta^2\over\theta_c^2}\right)
\left({1\over 3}-{\theta^2\over\theta_c^2}\right).
\label{Ufilter}
\end{equation}

The aperture mass is insensitive to the mass sheet degeneracy, and
therefore it provides an unambiguous splitting of the $E$ and $B$
modes. The drawback is that aperture mass is a much better estimate of
the small scale power than the large scale power.  It is obvious from
the function $J_4(k\theta_c)$ in Eq.(\ref{theomap}) which peaks at
$k\theta_c\sim 5$, and essentially, all scales larger than a fifth of
the largest survey scale remain unaccessible to $M_{\rm ap}$. The
large-scale part of the lensing signal is lost by $M_{\rm ap}$, while
the remaining small-scale fraction is difficult to interpret because
the strongly non-linear power is difficult to predict accurately
\citep{VW02}. 

It is therefore preferable to use the shear correlation function which
is a much deeper probe of the linear regime.  The $E$ and $B$ modes
can also be measured separately from the shear variance and the shear
correlation functions (Eq.\ref{theovariance}, \ref{theogg}).  However,
in contrast with the $M_{\rm ap}$ statistics, the separation of the
two modes can only be done up to an integration constant
\citep{CNPT01b}, which depends on the extrapolated signal outside the
measurement range, either at small or large scale.

An alternative, which does not require the knowledge of the signal
outside the measured range, is to use the aperture mass $B$ mode to
calibrate the shear correlation function $B$ mode.  An range
$\Delta\theta$ of angular scales where $\langle M_{\rm
ap}^2\left(\Delta\theta\right)\rangle_\perp\sim 0$ would guarantee that
the $B$ mode of the shear correlation function should be zero as well
(within the error bars), at angular scales $\sim 5 \Delta \theta$.
The practical implementation of this calibration scheme is discussed
in Section 5.

The $E$ and $B$ modes of the shear top-hat variance and correlation
function are accessible from the following shear correlation function
$\xi_+$ and $\xi_-$:

\begin{eqnarray}
\xi_+(r)&=&
=\langle \gamma_t(\theta)\gamma_t(\theta+r)\rangle+\langle \gamma_r(\theta)\gamma_r(\theta+r)
\rangle.\nonumber \\
\xi_-(r)&=&\langle \gamma_t(\theta)\gamma_t(\theta+r)\rangle -\langle \gamma_r(\theta)
\gamma_r(\theta+r)\rangle,
\label{xipm}
\end{eqnarray}

where $\gamma_t$ and $\gamma_r$ are the tangential and radial
projections of the shear on the local frame joining two galaxies
separated from a distance $r$.  Following \cite{CNPT01a}, we define

\begin{equation}
\xi'(r)=\xi_-(r)+4\int_r^\infty \frac{dr'}{r'} \xi_-(r')
	-12r^2 \int_r^\infty \frac{dr'}{r'^3}\xi_-(r').
\label{eqn:xipr}
\end{equation}
The $E$ and $B$ shear correlation functions are given by
\begin{equation}
\xi^E(r)=\frac{\xi_+(r)+\xi'(r)}{2}\ \ \ \ \ \ 
\xi^B(r)=\frac{\xi_+(r)-\xi'(r)}{2}.
\label{eqn:xieb}
\end{equation}
In this paper, we use Eq.(\ref{eqn:xieb}) as the cosmic shear signal for constraining
the cosmological model. The aperture mass (Eq.\ref{theomap}) is used only to
calibrate the $B$ mode of the shear correlation function, at scales where
$\langle M_{\rm ap}^2\rangle_\perp$ is measured to be zero.

\section{The VIRMOS data set, redshift distribution and statistical estimators}

We use the observations carried out within the VIRMOS-DESCART project
\footnote{http://terapix.iap.fr/DESCART} by the VIRMOS
\footnote{http://www.astrsp-mrs.fr} imaging and spectroscopic survey
\citep{OLFetal04}.
The data cover an effective area of 8.5 sq.deg. (12 sq.deg.
before masking) in the I-band, with a
limiting magnitude $I_{AB}=24.5$. Technical details of the data set are
given in \cite{VW01} and \cite{HJMCC03}. We applied a bright magnitude cut
at $I_{AB}=21$ in order to better exclude  foreground objects from the
source galaxies.

As we already  described in \cite{VW02}, we use an estimate of the source
redshift distribution based on photometric redshifts measured in
joint VLT and HDF data. The source redshift distribution is parametrised as

\begin{equation}
n(z)={\beta\over z_s \ \Gamma\left({1+\alpha\over \beta}\right)} 
\left({z\over
z_s}\right)^\alpha \exp\left[-\left({z\over z_s}\right)^\beta\right],
\label{zsource}
\end{equation}
with $\alpha=2$, $\beta=1.2$ and $z_s=0.44$. For these values of
$\alpha$ and $\beta$, the mean redshift is $\bar z_s \approx 2.1\ z_s$
and the median redshift is $\approx 1.9\ z_s$. These values are given
by the first VVDS data sets, and provide an accurate calibration of
the source redshift distribution used in VIRMOS-Descart. It agrees
nicely with spectroscopic redshift measurements in the VVDS survey
\citep[a preliminary release can be found in][]{OLFetal03}.  The
uncertainty in the source redshift distribution is discussed in
Section 5.

The galaxy shapes are measured and analysed with IMCAT, which is
described in \cite{KSB95} to which we refer for technical
details. This technique allows us to measure, for each galaxy $i$, an
ellipticity $\eg(\thetag_i)=(e_1,e_2)$ and a weight $w_i$ for each
galaxy and star in the data. The ellipticity is given by a weighted
second order moment of the light distribution $f(\thetag)$ of the
galaxy
%
which is an unbiased estimate of the shear $\gammag(\thetag_i)$.
The signals measured from the galaxy shapes are the binned tangential and
radial shear correlation functions.  These are given by a sum over
galaxy pairs $(\thetag_i,\thetag_j)$

\begin{eqnarray}
\xi_{tt}(r)&=&{\displaystyle\sum_{i,j} w_i w_j e_t(\thetag_i)\cdot e_t(\thetag_j)
\over \displaystyle\sum_{i,j} w_i w_j}\nonumber \\
\xi_{rr}(r)&=&{\displaystyle\sum_{i,j} w_i w_j e_r(\thetag_i)\cdot e_r(\thetag_j)
\over \displaystyle\sum_{i,j} w_i w_j},
\label{corrfct}
\end{eqnarray}
where $r=|\thetag_i-\thetag_j|$, and $(e_t, e_r)$ are the tangential
and radial ellipticities defined in the frame of the line connecting a
pair of galaxies.  The weights $w_i$ are computed from
the intrinsic ellipticity variance $\sigma_e^2$ and the r.m.s. of the
ellipticity PSF correction $\sigma_\epsilon^2$
\citep[for the details, see Section 3 in ][]{VW02}.
\footnote{In practice, $r$ is the averaged angular distance computed
inside very tiny bin size ($\approx $ 3 pixels only). Note that the
$r'$ value of the shear statistics is different and correspond to the
upper limit: $r<r'$ }

\section{Updated analysis of the VIRMOS-Descart data}

\subsection{Previous sources of $B$-mode}

\cite{VW02} discussed the presence of residual systematics, and
attributed the $B$ mode to an imperfect correction for PSF anisotropy.
The presence of this $B$ mode limited the precision with which the
lensing signal could be determined. The $B$-mode
was comparable to the $1\sigma$ error. As it was
not clear how to reduce it and since there was also no obvious
way to quantify the contamination on the $E$ mode, \cite{VW02} did not
consider any further improvement. Rather, they included the
systematics in the final error budget, leading to a slightly
biased normalisation of the mass power spectrum towards high values.

\cite{HOEK04} studied the effects of PSF anisotropy using CFHT
observations of fields with a high number density of stars. This study
revealed that the second order polynomial model for the PSF anisotropy
used in \cite{VW02} does not accurately describe the spatial variation
of the systematics. \cite{HOEK04} showed that there are rapid changes
in the PSF anisotropy at the edges of the CFH12k mosaic. In the actual
VIRMOS images, the small scale PSF correction is made difficult by the
lack of stars. The mean star separation at high galactic latitude is
slightly less than $2'$. Therefore the choice of the correct PSF
correction model must play an important role for the small scale
cosmic shear signal \citep{HOEK04}.

In order to estimate the effect of model choices, we tested various
PSF models for the spatial variation of the PSF anisotropy. The
ellipticities of the observed stars are modeled using
generic functions:
 
\begin{equation}
e_{*}^{\rm model}(x,y) = \frac{\sum_{l,m} \ a_{lm} \ f_{lm}(x,y)} 
{1+\sum_{l,m} \ b_{lm} \ f_{lm}(x,y)},
\end{equation}
where the $f_{lm}(x,y)$ are the generic formal polynomial function
terms listed in Table \ref{psftable} and the $a_{lm}$ and $b_{lm}$ are
their coefficients, as derived from a least squares fit of the model
to the sample of stars selected from the data. Table~\ref{psftable}
lists six of the PSF models that were considered. Models $1-5$ are
simple polynomials (i.e., $b_{lm}=0$), with higher orders in $y$
because the CFH12k chips are longer in that direction and because the
rapid change in PSF anisotropy is a strong function of $y$. Model 6 is
a ``tailor'' made model, based on the study of dense star fields
\citep{HOEK04}.

\begin{figure}
\centerline{
\psfig{figure=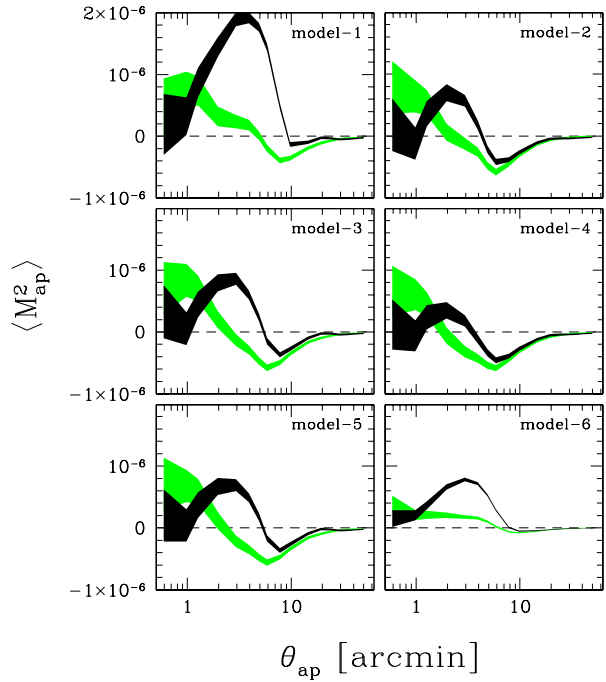,width=9.5cm}
}
\caption{\label{stars_corrfct.ps} 
The $E$ (black area) and $B$-modes (light/green area) aperture mass
statistics computed from the residual shapes of the stars after
correcting using on of the six PSF models discussed in the text. The
areas indicate the $1\sigma$ regions around the mean.}
\end{figure}

For each model, we measured the ellipticity correlation functions from
the stars (after correction) and computed the corresponding $E$ and
$B$ mode aperture mass statistics. The results are presented in
Figure~\ref{stars_corrfct.ps}. We find significant deviations from
$E=0$ and $B=0$ at small scale ($<10'$) for all models. Note that
these results do not tell us about the amount of residual systematics
in the galaxy signal, although they can be used to make an estimate
\citep[see][]{HOEK04}. 

In the absence of more detailed information,
Figure~\ref{stars_corrfct.ps} provides a very useful way to choose
which of the PSF correction models works best. The rational function
(model 6) shows the smallest contamination on all scales. In
particular the deviation from zero starts at a smaller scale than the
other models. Figure~\ref{stars_corrfct.ps} demonstrates there is
still room left for an even better correction at small scales, but the
lack of suitable stars makes this a very challenging issue.

\begin{table}
\caption{Different Point Spread Function parametric models used to
correct the star anisotropy in the VIRMOS data. The table lists the
basis of functions of the models. Model 6 is a rational function,
the other models have $b_{lm}=0$.}
\label{psftable}
\begin{center}
\begin{tabular}{|c|c|}
\hline
model \# & Correction Model\\
\hline
1 & $(1,x,y,x^2,xy,y^2)$ \\
2 & $(1,x,y,x^2,xy,y^2,x^3,y^3,y^4)$ \\
3 & $(1,x,y,x^2,xy,y^2,y^3)$ \\
4 & $(1,x,y,x^2,xy,y^2,y^3,y^4,y^5,x^3,x^4)$ \\
5 & $(1,x,y,x^2,xy,y^2,x^3,y^3)$ \\
6 & $(1,x,y,x^2,xy,y^2,y^3,y^4)/(x,y)$\\
\hline
\end{tabular}
\end{center}
\end{table}

\begin{figure}
\centerline{
\psfig{figure=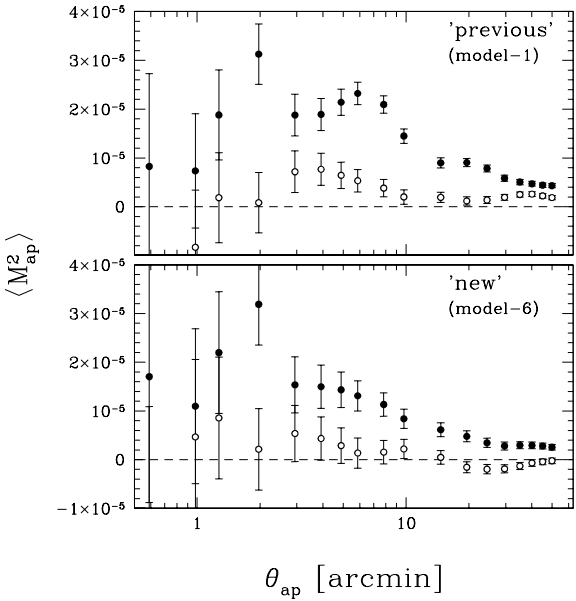,width=8.5cm}
}

\caption{\label{compare.ps} {\it top panel}: $E$ (filled points) and $B$
(open points) mode aperture mass measurements as measured from the
previous analysis. It uses model-1 (see Table \ref{psftable}) to
describe the PSF anisotropy with the wrong centroid estimate and no
sigma clipping on the shape parameters of the selected stars.
{\it bottom panel}: results using model~6 and the improved catalogues.
Because of a different object selection in the new version, we
ensured that we only used objects that are in common to produce this
Figure. The new catalogue contains more galaxies.}
\end{figure}

These new insights in the correction for PSF anisotropy have led to a
significant reduction in the level of systematics of the
VIRMOS-Descart cosmic shear signal, as we will demonstrate below.
Several other sources of the $B$ mode observed by \cite{VW02} were
identified as well. The systematics that have been corrected for in
the revised analysis presented here are:

\begin{enumerate}
\item The small scale $B$ mode was generated by an imperfect PSF correction model
as described above. As suggested in \cite{HOEK04} a proper correction for
PSF anisotropy lowered the $E$ mode signal on these scales.

\item The large scale constant $B$ mode was caused by an imperfect
calculation of the centroid of the objects. This problem was the most
difficult to find because it was caused by a subtle inconsistency
between the centroid estimate (based on SExtractor) and the shape
calculation (based on IMCAT). On one of the VIRMOS fields (F10),
SExtractor systematically computed a biased centroid which was the
cause of the large scale bias in shape measurement generating a $B$
mode. This problem was cured by measuring the centroids using aperture
filters instead of isophotal apertures.

\item The third important source of systematics arose from spurious
ellipticies for some of the selected stars. Some of the stars
survived the selection criteria based on a sigma clipping
applied to their {\it corrected}
ellipticity.  It turned out that some stars passed this selection
criteria, but their smear polarisability tensor was significantly 
different from the bulk of the stars \citep[for a definition
see ][]{KSB95}. This could happen when a star has a small very close
companion which would contaminate the derivative of the profile much
more than its ellipticity. The application of a $3-\sigma$ clipping
to all of the star parameters,
corrected ellipticity, smear polarisability and the stellar
part of the preseeing shear polarisability remove all the remaining
unwanted stars which contributed to the residual $B$ mode.
\end{enumerate}

Figure \ref{compare.ps} demonstrates how the results have changed from
the early VIRMOS analysis (top panel) and the current analysis
(bottom). The top panel shows clearly the small and large scale
residual $B$ mode as measured before and a ``bump'' in the $E$ mode at
$\sim 5$ arcminutes. The bottom panel shows the improvement after the
main residual systematics have been corrected for. It also shows that
the aperture mass is an effective statistic to identify and quantify
residual systematics.

\begin{figure*}
\centerline{
\psfig{figure=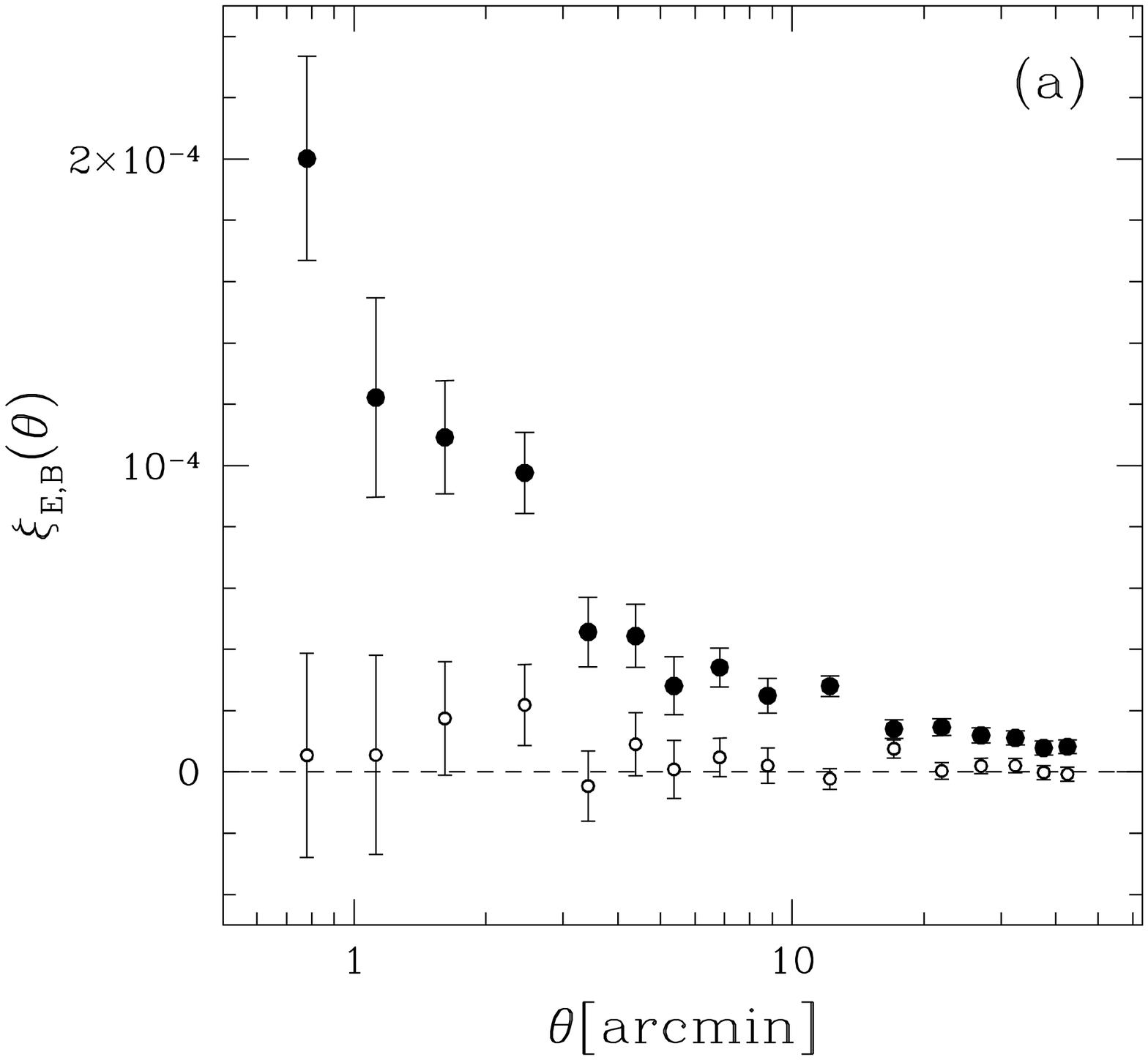,height=8cm}
\psfig{figure=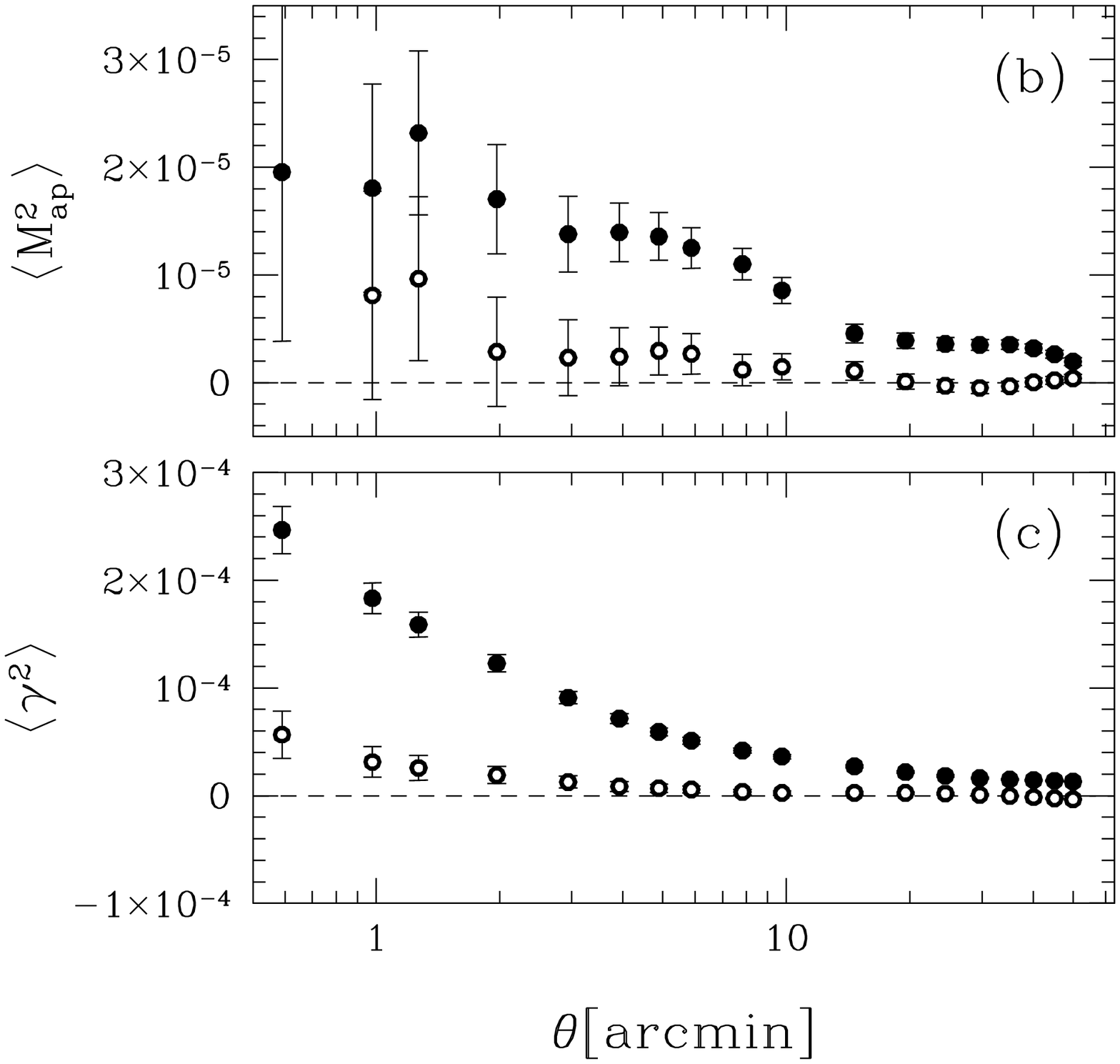,height=8cm}
}
\caption{\label{xi.ps} {\it Panel (a)}: $E$ and $B$ mode shear correlation
function (filled and open points, respectively) as measured from the
VIRMOS-Descart data; {\it panel (b)}: $E$ and $B$ mode measurements
for the aperture mass statistic; {\it panel (c)}: as before but now
for the top-hat variance. For all three statistics we observe
the signal is consistent with no $B$-mode on large scales, and a 
hint of minor contamination on small scales.}
\end{figure*}

\subsection{Residual systematics}

The results of the new analysis of the VIRMOS-Descart are presented in
Figure~\ref{xi.ps}. To derive these measurements, all sources of
systematics discussed in the previous section have been accounted for,
and model~6 was used to characterize the spatial variation of the PSF
anisotropy. Figure~\ref{xi.ps}a shows the resulting ellipticity
correlation function as a function of angular scale. The measurements
are tabulated in Table~\ref{xitable}. As explained in Section 2,
in order to separate the signal into an $E$ and $B$ mode for
the top-hat shear variance and shear correlation function,
we have to define a zero-point for the $B$
mode. Based on the results for the aperture mass statistic (panel~b)
the correlation functions are scaled such that the mean $B$ mode is
zero on scales $2-10'$ according to the vanishing $B$ mode for the
aperture mass at $\theta=10-50'$. The error bars reflect only the
statistical error bars. There is no cosmic variance on the $B$ mode,
and consequently the statistical error is an unbiased estimate of the
$B$ mode noise.

Panels~(b) and~(c) show the aperture mass $\langle M_{\rm
ap}^2\rangle$ and the shear top-hat variance $\langle\gamma^2\rangle$,
respectively. The measurements of both statistics are listed in
Table~\ref{varmaptable}. The small scale $\langle M_{\rm ap}^2\rangle$
$B$ mode is consistent with no signal (i.e., no systematics). However,
one has to bear in mind that different PSF correction models produce
fluctuations up to $10\%$ in the amplitude of the aperture mass below
$1'$. On the other hand, the aperture mass signal is robust against
changes in the adopted PSF correction model for the range of scales
$10-50'$ which corresponds to a physical angular scale range of
$2-10'$. This means that the smallest scales in the aperture mass
statistics give robust cosmological signal only if the PSF variation
is properly removed.

The amount of residual systematics left in the signal due to imperfect
PSF correction can be estimated from the correlation between the
uncorrected stars and the corrected galaxies. Such an estimator was
defined in \cite{Hey03}:

\begin{equation}
\xi_{\rm SYS}={\langle e^\star\gamma\rangle^2\over \langle e^\star e^\star\rangle}.
\end{equation}
where $e^\star$ is the star ellipticity before PSF correction, and
$\gamma$ is the shear estimate of the galaxies. $\xi_{\rm SYS}$ is
renormalised by the star ellipticity correlation function $\langle
e^\star e^\star\rangle$, which makes it directly comparable to the
signal $\langle\gamma(r)\gamma(\theta+r)\rangle$, and not dependent on
$|e^\star |$.  

\begin{figure}
\centerline{
\psfig{figure=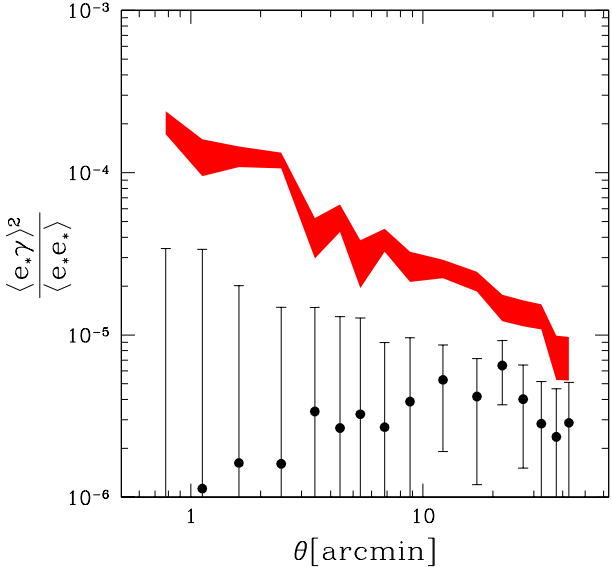,width=8.5cm}
}
\caption{\label{xibias.ps} The shades (red) area corresponds to the
$1\sigma$ area around the mean shear correlation function
$\langle\gamma(r)\gamma(\theta+r)\rangle$ measured from the
VIRMOS-Descart data. The points with error bars show the residual
systematics correlation function $\xi_{\rm SYS}$ as defined in the
text. The indicated error bars are those of the shear correlation
function, and are displayed to show the level of systematics
with respect to the $1-\sigma$ statistical error.}
\end{figure}

Figure \ref{xibias.ps} shows $\xi_{\rm SYS}$ for the VIRMOS data. The
residual PSF contamination is consistent with no signal except maybe
for one point around $\theta \simeq 22 {\rm arcmin}$, which is at
2.4$\sigma$.  It demonstrates that
the anisotropy of the PSF has been almost completely removed.  
However, this plot provides no information about the accuracy of the
isotropic PSF correction (the ``pre-seeing'' shear polarisability). As
outlined in \cite{HS03}, the accuracy of the isotropic correction of
the PSF is still somewhat uncertain, essentially because there is yet no
direct way to obtain a perfect calibration of the shear amplitude to
compensate for the PSF smearing. One should emphasize that \cite{E01} and
\cite{BRCE01}
used simulated images to demonstrate that the shape
measurement method used in this paper \citep{KSB95} is accurate to
$5\%$ or better at recovering the correct shear amplitude. The worst calibration
was obtained for the most ellipticial galaxies. Hence, we expect an accuracy
due to isotropic calibration smaller than $5\%$ in this study, which is
still smaller than the statistical error. The
PSF analysis technique based on reducing the $B$ mode
amplitude shows how it is possible to control the PSF anisotropy
correction, however the isotropic correction still has to be checked
using simulated images.

\begin{table}
\caption{$E$ and $B$ modes of the shear correlation function. The error is statistical only
(see Section 6 for the covariance error).
}
\label{xitable}
\begin{center}
\begin{tabular}{|c|c|c|c|}
\hline
$\theta$ & $\xi_E$ & $\xi_B$ & $\delta\xi$ \\
\hline
     0.780000 &  0.000200166 &  5.43400e-06 &  3.33000e-05 \\
      1.12000 &  0.000122166 &  5.53400e-06 &  3.25000e-05 \\
      1.61000 &  0.000109166 &  1.74340e-05 &  1.85000e-05 \\
      2.45000 &  9.75660e-05 &  2.18340e-05 &  1.32000e-05 \\
      3.42000 &  4.56660e-05 & -4.60600e-06 &  1.14000e-05 \\
      4.39000 &  4.43660e-05 &  9.03400e-06 &  1.03000e-05 \\
      5.37000 &  2.80660e-05 &  8.14000e-07 &  9.46000e-06 \\
      6.83000 &  3.40660e-05 &  4.73400e-06 &  6.29000e-06 \\
      8.79000 &  2.48660e-05 &  2.03400e-06 &  5.71000e-06 \\
      12.2000 &  2.79660e-05 & -2.24600e-06 &  3.38000e-06 \\
      17.0800 &  1.39860e-05 &  7.53400e-06 &  2.98000e-06 \\
      21.9600 &  1.46060e-05 &  3.04000e-07 &  2.76000e-06 \\
      27.0100 &  1.18960e-05 &  1.93400e-06 &  2.51000e-06 \\
      32.3300 &  1.10960e-05 &  2.03400e-06 &  2.32000e-06 \\
      37.5400 &  7.80100e-06 & -2.06000e-07 &  2.31000e-06 \\
      42.5400 &  8.22120e-06 & -7.46000e-07 &  2.23000e-06 \\
\hline
\end{tabular}
\end{center}
\end{table}

\begin{table*}
\caption{Values of the aperture mass and shear top-hat variance as a function of scale.
The $B$ mode is rescaled to zero for the later, and the error is statistical only.
}
\label{varmaptable}
\begin{center}
\begin{tabular}{|c|c|c|c|c|c|c|}
\hline
$\theta$ & $\langle M_{\rm ap}^2\rangle$ & $\langle M_{\rm ap}^2\rangle_\perp$ & $\delta\langle M_{\rm ap}^2\rangle$ & $\langle\gamma_E^2\rangle$ & $\langle\gamma_B^2\rangle$ &$\delta\langle\gamma^2\rangle$\\
\hline
 0.59 & 0.1956E-04 & -.2560E-04 & 0.1571E-04 & 0.000246658 &  5.66021e-05 &  2.18700e-05 \\
 0.98 & 0.1806E-04 & 0.8109E-05 & 0.9666E-05 & 0.000183258 &  3.11921e-05 &  1.41900e-05 \\
 1.27 & 0.2318E-04 & 0.9647E-05 & 0.7599E-05 & 0.000158658 &  2.56821e-05 &  1.14700e-05 \\
 1.97 & 0.1703E-04 & 0.2864E-05 & 0.5083E-05 & 0.000122858 &  1.90721e-05 &  8.00500e-06 \\
 2.94 & 0.1378E-04 & 0.2319E-05 & 0.3512E-05 & 9.08879e-05 &  1.26021e-05 &  5.69900e-06 \\
 3.92 & 0.1395E-04 & 0.2413E-05 & 0.2704E-05 & 7.15379e-05 &  8.56213e-06 &  4.45700e-06 \\
 4.89 & 0.1357E-04 & 0.2952E-05 & 0.2217E-05 & 5.92279e-05 &  6.85213e-06 &  3.68800e-06 \\
 5.87 & 0.1250E-04 & 0.2675E-05 & 0.1884E-05 & 5.09079e-05 &  5.65212e-06 &  3.15500e-06 \\
 7.82 & 0.1100E-04 & 0.1198E-05 & 0.1465E-05 & 4.18679e-05 &  3.34212e-06 &  2.47200e-06 \\
 9.76 & 0.8562E-05 & 0.1469E-05 & 0.1210E-05 & 3.63379e-05 &  2.52213e-06 &  2.04900e-06 \\
14.65 & 0.4570E-05 & 0.1084E-05 & 0.8642E-06 & 2.72479e-05 &  2.59212e-06 &  1.46700e-06 \\
19.52 & 0.3899E-05 & 0.1008E-06 & 0.6924E-06 & 2.18579e-05 &  2.45213e-06 &  1.17100e-06 \\
24.41 & 0.3594E-05 & -.2831E-06 & 0.5858E-06 & 1.83569e-05 &  1.92213e-06 &  9.88100e-07 \\
29.62 & 0.3503E-05 & -.4899E-06 & 0.5065E-06 & 1.61539e-05 &  7.78125e-07 &  8.55300e-07 \\
35.05 & 0.3537E-05 & -.3382E-06 & 0.4477E-06 & 1.50329e-05 & -3.31876e-07 &  7.56800e-07 \\
40.05 & 0.3197E-05 & 0.4226E-07 & 0.4088E-06 & 1.43569e-05 & -1.40988e-06 &  6.89600e-07 \\
45.05 & 0.2635E-05 & 0.2219E-06 & 0.3804E-06 & 1.37509e-05 & -2.50788e-06 &  6.38200e-07 \\
49.95 & 0.1956E-05 & 0.3912E-06 & 0.3599E-06 & 1.30189e-05 & -3.49488e-06 &  5.99200e-07 \\
\hline
\end{tabular}
\end{center}
\end{table*}

\section{Parameter estimation}

\subsection{Likelihood calculation}

Following \cite{VW02}, we investigate a four dimensional parameter
space defined by the mean matter density $\Omega_M$, the normalisation
of the power spectrum $\sigma_8$, the shape parameter $\Gamma$ and the
redshift of the sources parameterized by $z_s$ (see
Eq.(\ref{zsource})). We adopt a flat cosmology, unless stated
otherwise.  The default priors are taken to be $\Omega_M \in [0.1,1]$,
$\sigma_8 \in [0.5,1.3]$, $\Gamma \in [0.1,0.5]$ and $z_s \in
[0.38,0.48]$. The latter corresponds to a mean redshift between $0.8$
and $1.0$, in agreement with \cite{OLFetal04}.

The data vector $\xi_i$ is the shear correlation function $\xi_E$ as a
function of scale, as listed in Table \ref{xitable}. This table also
shows that the residual systematic $\xi_B$ is negligible. If we take
$m_i(\Omega_M,\sigma_8,\Gamma,z_s)$ to be the model prediction, the
likelihood function of the data is given by

\begin{equation}
{\cal L}={1\over (2\pi)^n|\Cg|^{1/2}} \exp\left[(\xi_i-m_i)\Cg^{-1}(\xi_i-m_i)^T\right],
\end{equation}
where $n=16$ is the number of angular scale bins and $\Cg$ is the $16\times 16$
covariance matrix,

\begin{equation}
C_{ij}=\langle (\xi_i-m_i)^T(\xi_j-m_j)\rangle.
\end{equation}
$\Cg$ can be decomposed as $\Cg=\Cg_n+\Cg_s$, where $\Cg_n$ is the
statistical noise and $\Cg_s$ the cosmic variance covariance matrix.
$\Cg_n$ is diagonal and given in Table \ref{xitable} (last columns).
The matrix $\Cg_s$ is computed according to \cite{S02}, assuming an
effective survey area of $8.5$ square degrees, a number density of
galaxies $n_{gal}=15 {\rm \ per\ arcmin}^2$, and an intrinsic
ellipticity dispersion $\sigma_e=0.44$. These numbers can be found in
\cite{VW02}. The cosmic variance is computed assuming a Gaussian
statistic, but as shown in \cite{VW02}, this is not an issue in our
case: the non gaussian contribution appears only at small scale, where
the noise is still dominated by the statistical error. We need a
fiducial model to compute the matrix components, which is taken as the
best fit model following the WMAP results \citep{Sp03}. It corresponds
to the choice $\Omega_M=0.3$, $\Lambda=0.7$, $\sigma_8=0.85$,
$\Gamma=0.21$ (the reduced Hubble constant is $h$=0.7). We take a mean
source redshift for the fiducial model of $0.9$, which corresponds to
$z_s=0.43$.  Table \ref{Ctable} gives the full covariance matrix $\Cg$
used in this paper. The pure cosmic variance part of the matrix
($\Cg_s$) can easily be extracted by subtracting the statistical noise
contribution given in Table \ref{xitable}. The $B$ mode is calibrated
by marginalizing around $B=0$ within the 1$\sigma$ interval.

\begin{figure}
\centerline{
\psfig{figure=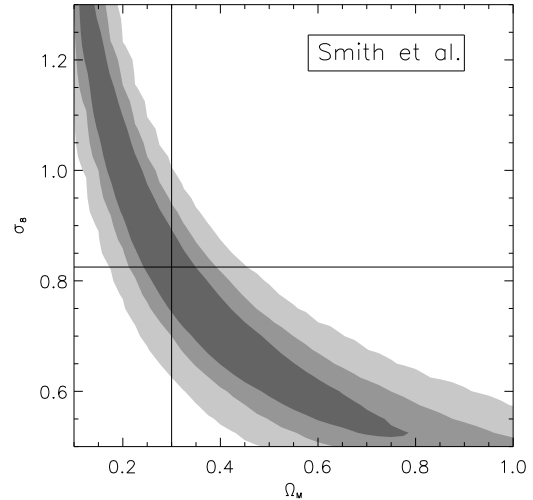,height=7cm}
}
\caption{\label{PD_smith.ps} Constraints on $\sigma_8$ and $\Omega_M$
for a flat CDM Universe, adopting the non-linear prediction
given by \cite{S03}. The contours correspond to the $0.68$, $0.95$
and $0.999$ confidence levels.  The solid dark vertical line indicates
$\Omega=0.3$ and the horizontal line indicates the maximum of the
likelihood for $\sigma_8$ at $\Omega=0.3$.  The models are
marginalised over $\Gamma\in[0.1,0.3]$ and $z_s\in [0.4,0.48]$.}
\end{figure}

\begin{figure}
\centerline{
\psfig{figure=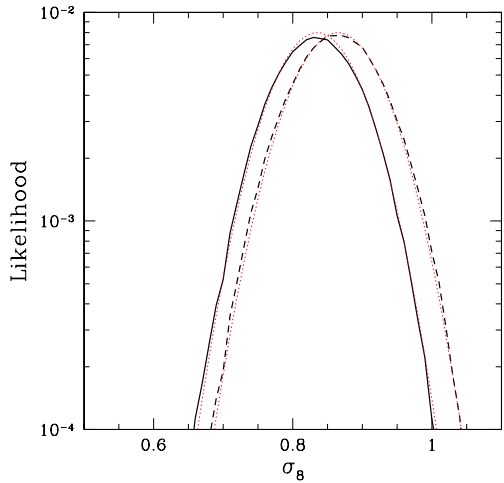,height=7cm} }
\caption{\label{fig6.eps} $\sigma_8$ probability distribution
functions for $\Omega_M=0.3$. Dashed curve is for \cite{PD96} (high
normalisation), and the solid curve for \cite{S03} (low
normalisation). The superimposed dotted lines indicate a Gaussian fit to
the likelihood function.}
\end{figure}

\begin{table*}
\caption{Full covariance matrix $\Cg$, as discussed in the text,
Section 6. Units are $10^{-10}$.  The scales $\theta_i$ correspond to
those given in Table \ref{xitable}.  }
\label{Ctable}
\begin{center}
\begin{tabular}{|c|c|c|c|c|c|c|c|c|c|c|c|c|c|c|c|c|}
\hline
 & $\theta_1$ &$\theta_2$ &$\theta_3$ &$\theta_4$ &$\theta_5$ &$\theta_6$ &$\theta_7$ &$\theta_8$ &$\theta_9$ &$\theta_{10}$ &$\theta_{11}$ &$\theta_{12}$ &$\theta_{13}$ &$\theta_{14}$ &$\theta_{15}$ &$\theta_{16}$ \\          
\hline
$\theta_1$ &     14.48 &       2.15 &       1.85 &       1.55 &       1.33 &       1.20 &       1.10 &       1.01 &       0.92 &       0.80 &       0.70 &       0.62 &       0.56 &       0.49 &       0.44 &       0.40 \\
$\theta_2$ &      2.16 &      13.57 &       1.87 &       1.56 &       1.34 &       1.20 &       1.10 &       1.01 &       0.92 &       0.81 &       0.70 &       0.62 &       0.56 &       0.49 &       0.44 &       0.40 \\
$\theta_3$ &      1.86 &       1.88 &       5.57 &       1.57 &       1.35 &       1.21 &       1.11 &       1.01 &       0.92 &       0.81 &       0.70 &       0.62 &       0.56 &       0.49 &       0.44 &       0.40 \\
$\theta_4$ &      1.56 &       1.57 &       1.58 &       3.46 &       1.38 &       1.22 &       1.12 &       1.02 &       0.92 &       0.81 &       0.70 &       0.62 &       0.56 &       0.49 &       0.44 &       0.40 \\
$\theta_5$ &      1.34 &       1.35 &       1.36 &       1.38 &       2.78 &       1.25 &       1.15 &       1.04 &       0.92 &       0.81 &       0.70 &       0.62 &       0.56 &       0.49 &       0.44 &       0.40 \\
$\theta_6$ &      1.21 &       1.21 &       1.21 &       1.23 &       1.26 &       2.41 &       1.16 &       1.04 &       0.93 &       0.81 &       0.70 &       0.62 &       0.56 &       0.49 &       0.44 &       0.40 \\
$\theta_7$ &      1.10 &       1.10 &       1.11 &       1.12 &       1.15 &       1.17 &       2.14 &       1.05 &       0.93 &       0.81 &       0.70 &       0.61 &       0.56 &       0.49 &       0.44 &       0.40 \\
$\theta_8$ &      1.01 &       1.01 &       1.03 &       1.03 &       1.04 &       1.04 &       1.05 &       1.50 &       0.94 &       0.81 &       0.69 &       0.62 &       0.56 &       0.49 &       0.44 &       0.40 \\
$\theta_9$ &      0.92 &       0.92 &       0.92 &       0.93 &       0.92 &       0.93 &       0.93 &       0.95 &       1.32 &       0.82 &       0.70 &       0.62 &       0.56 &       0.49 &       0.44 &       0.40 \\
$\theta_0$ &      0.80 &       0.80 &       0.81 &       0.81 &       0.81 &       0.81 &       0.81 &       0.81 &       0.82 &       0.96 &       0.70 &       0.62 &       0.55 &       0.48 &       0.44 &       0.40 \\
$\theta_{11}$ &      0.70 &       0.70 &       0.70 &       0.70 &       0.70 &       0.70 &       0.70 &       0.69 &       0.70 &       0.71 &       0.82 &       0.62 &       0.55 &       0.51 &       0.44 &       0.41 \\
$\theta_{12}$ &      0.62 &       0.62 &       0.62 &       0.62 &       0.62 &       0.62 &       0.62 &       0.62 &       0.63 &       0.62 &       0.62 &       0.72 &       0.55 &       0.48 &       0.44 &       0.39 \\
$\theta_{13}$ &      0.56 &       0.56 &       0.56 &       0.56 &       0.56 &       0.56 &       0.56 &       0.56 &       0.56 &       0.56 &       0.55 &       0.55 &       0.62 &       0.49 &       0.44 &       0.39 \\
$\theta_{14}$ &      0.49 &       0.49 &       0.49 &       0.49 &       0.49 &       0.49 &       0.49 &       0.50 &       0.50 &       0.49 &       0.51 &       0.48 &       0.49 &       0.56 &       0.44 &       0.39 \\
$\theta_{15}$ &      0.44 &       0.44 &       0.44 &       0.44 &       0.44 &       0.44 &       0.44 &       0.44 &       0.44 &       0.44 &       0.44 &       0.44 &       0.44 &       0.44 &       0.50 &       0.39 \\
$\theta_{16}$ &      0.40 &       0.40 &       0.40 &       0.40 &       0.40 &       0.40 &       0.40 &       0.40 &       0.40 &       0.40 &       0.41 &       0.39 &       0.40 &       0.40 &       0.40 &       0.46 \\
\hline
\end{tabular}
\end{center}
\end{table*}
The results of the maximum likelihood analysis are presented in Figure
\ref{PD_smith.ps}, which shows the constraints in the
$\Omega_M$-$\sigma_8$ plan for the non-linear prediction schemes
described in \cite{S03}. We note that we find similar results for the
\cite{PD96} model, although the best fit cosmologies differ somewhat
(e.g., see Figure~\ref{fig6.eps}). We marginalised over
$\Gamma\in[0.1,0.3]$ and $z_s\in [0.4,0.48]$, assuming a flat
probability distribution in these intervals. For the \cite{S03}
model we find 

\begin{equation}
\sigma_8=(0.83\pm 0.07)\left(\Omega_m\over 0.3\right)^{-0.49}.
\end{equation}

\noindent For the \cite{PD96} model we find a somewhat higher
normalisation, given by

\begin{equation}
\sigma_8=(0.86\pm 0.08)\left(\Omega_m\over 0.3\right)^{-0.49}.
\end{equation}

The difference between these two non-linear models corresponds to a
4\% discrepancy, and it shows the limitation in the cosmic shear
signal interpretation due to non-linear modeling. As the \cite{S03}
is the most recent, and more extensive analysis of the non-linear
power spectrum, we prefer the results of this model. The updated
results are also in excellent agreement with results obtained
from the Red-sequence Cluster Survey \citep{HOEK02b}, who
found $\sigma_8=0.86^{+0.04}_{-0.05}$ for $\Omega_m=0.3$. They
are also in agreement with the revisited CTIO lensing
survey \citep{JJ04}.

Figure \ref{fig6.eps} shows the probability distribution functions for
$\sigma_8$ for the two models, adopting $\Omega_m=0.3$. Despite the
different maximum likelihood shifts between the models, the curves are
well fitted by a gaussian distribution as shown by the dashed line.

It is interesting to note that the new PSF anisotropy correction drops
the resulting $\sigma_8$ only by 10\% as compared to \cite{VW01}, much
less than what would be expected from the change in amplitude in the
lensing signal. However, in \cite{VW01}, the source redshift
distribution was given by the model described in \cite{Wil01} (their
figure 4), which turns out to give too much weight to high redshift
galaxies compared to what we know about the VIRMOS galaxy sample
today \citep{OLFetal03,OLFetal04}.
. Hence the residual systematics in \cite{VW01} were partly
compensated by this biased redshift distribution model. For the
present analysis we use more recent results from a comparison with
photometric and spectroscopic redshifts \citep{OLFetal03,OLFetal04}. This
changes the redshift distribution and significantly reduces the 
associated uncertainties.

\subsection{Combination with WMAP}

Figure~\ref{PD_smith.ps} indicates that a high density Universe
$\Omega_M=1$ is excluded at the 2-$\sigma$ level. This result,
however, depends on the prior used for the shape parameter
$\Gamma\sim\Omega_M h$ (which is $\Gamma\in [0.1,0.3]$), and therefore
on the Hubble constant (and the baryon abundance to a less
extent). Using weak lensing data alone, it is still possible to
accommodate $\Omega_M\simeq 1$, provided $\Gamma$ is at least as large
as $0.5$, and we cannot exclude the high density ($\Omega_M > 0.9$)
alternative to the concordance model, proposed by \cite{Bl03}. The
latter model only works if the Hubble constant is low ($\sim 0.5$,
which is not in agreement with \cite{F01}) and if the Universe
expansion rate is not accelerating as suggested by the type Ia supernovae
results \citep{R98,P99}.

From the cosmic shear point of view, such a model would require a low
normalisation $\sigma_8\simeq 0.4$, because for higher values of
$\Gamma$, the banana shape contour in the $\Omega_M-\sigma_8$ plane
(Figure \ref{PD_smith.ps}) is tilted clockwise. For example, the
combination of $\Gamma=0.5$ and $\sigma_8=0.5$ is excluded at more
than 3-$\sigma$. This is inconsistent with WMAP \citep{Sp03} which
gives $\sigma_8=0.9\pm 0.1$ for a flat and adiabatic Cold Dark Matter
model assuming a power law primordial power spectrum. 

\begin{figure}
\centerline{ 
\psfig{figure=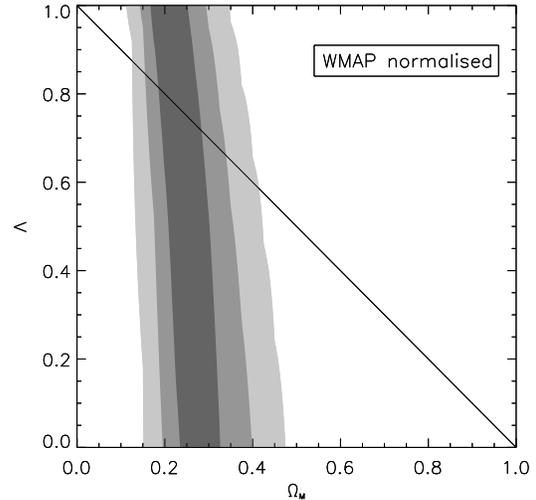,height=7cm} }
\caption{\label{omega_lambda.ps} VIRMOS-Descart data constraints in
the $\Lambda-\Omega_M$ plane.  The models are normalised to the WMAP
results ($\sigma_8=0.9\pm 0.1$), and we assumed the priors
$\Gamma=[0.1,0.5]$ and $z_s=[0.4,0.48]$.  }
\end{figure}

Combined with our cosmic shear results, this leads to a rejection of
the very low and high density universes, with a preferred value around
$\Omega_M \simeq 0.3$, as indicated by Figure~\ref{omega_lambda.ps}.
This result also demonstrates that CMB and cosmic shear experiments
are complementary, enabling one to break the $\Omega_M-\sigma_8$
degeneracy without external data sets \citep{C03,T04}. Although one
can modify the matter content of the Universe to try to account for a
low $\sigma_8$ at high $\Omega_M$ \citep{Bl03}, the next generation of
weak lensing survey will certainly give an unambiguous constraint on
$\sigma_8$ at low redshift regardless the value of $\Omega_M$.

\section{Conclusion}

We have presented a revised analysis of the VIRMOS-Descart cosmic
shear data, focussing on how to deal with residual systematics. The
main result of this paper is a $10\%$ smaller mass power spectrum
normalisation compared to our previous estimate. We find
$\sigma_8=0.83\pm 0.07$ (for the preferred non-linear power spectrum
as described by \cite{S03}), which is at the 1-$\sigma$ bottom end of
the measurement given in \cite{VW02}, where the systematics
discussed in this study were still present. Our better understanding of
systematics allows us to use weaker cosmic shear signal at higher
angular scale and therefore to be less sensitive to unknown
properties of the non-linear power spectrum.

In this work, various sources of systematics, which were previously
ignored, were identified and the amplitude of the residual $B$ mode
from the aperture mass was used to significantly improve the
cosmological lensing signal. The choice of a better model to describe
the spatial variation of PSF anisotropy at small scale proved to be
critical for an unbiased measurement. We demonstrated that small
differences between correction schemes can have a significant impact
on the small scale signal. Fortunately, the residual systematics in
the aperture mass of the stars provide an unambiguous method to test
for residual systematics by selecting the least biased model. 

On large scales, a few outliers in the star catalogue used to correct
for the effects of the PSF can introduce a bias even if their
ellipticities appear reasonable. It is therefore important to reject
{\it spurious} stars not only from their corrected ellipticity, but
using extended criteria such as the smear polarisability and the
preseeing shear polarisability. We also found that a large angular
scale bias can be introduced from the use of galaxy centroids based on
isophotal cuts. 

There is a fundamental limit to the accuracy of the correction at
small scales, which is determined by the sampling of the PSF variation
across the CCDs. Further high precision cosmic shear studies will
require the observation of dense star fields to improve the correction
\cite[see ][]{HOEK04}. Fortunately, the large scale signal ($> 10'$) 
appears unaffected by the choice of the PSF anisotropy model.

The use of the shear correlation function is much less sensitive to
systematics than the shear top-hat variance or even the aperture mass
variance. This is most critical for the aperture mass which is
sensitive to very small scales. On the other hand, the aperture mass
$B$ mode is a crucial test of the residual systematics, which has to
be used to calibrate the PSF correction and the $B$ mode of the shear
correlation function.

The prediction of the non-linear power spectrum results in values for
$\sigma_8$ that differ by a few percent for different models.  The
model presented by \cite{S03} always gives a smaller normalisation
than \cite{PD96}. Currently, the former provides the preferred model.
We also note that these differences are not yet an important issue for
surveys of the size we considered here. However, it will be
a challenging task for the forthcoming lensing surveys to improve
the non-linear predictions to the required level of precision.

It is still possible to accommodate a large value for $\Omega_m$
from cosmic shear measurements alone. However, the
combination of the cosmic shear signal with measurements of the Cosmic
Microwave Background provides strong evidence for a low density
universe without adding external data sets. To derive these
conclusions we do have to assume a Cold Dark Matter scenario with a
primordial power law power spectrum. Such assumptions can be
relaxed with the next generations of lensing surveys.

{\acknowledgements 
We thank Karim Benabed, Francis Bernardeau, Emmanuel Bertin, Heny McCracken,
Thomas Erben, Alexandre R\'efr\'egier, Peter Schneider, Elisabetta
Semboloni, and Ismael Tereno for useful discussions.  This
work was supported by the TMR Network ``Gravitational Lensing: New
Constraints on Cosmology and the Distribution of Dark Matter'' of the
EC under contract No. ERBFMRX-CT97-0172. We thank the TERAPIX data
center for providing its facilities for the data reduction of CFH12K
images.  }

\end{document}